\documentstyle[11pt]{article}
\newlength{\bredde}
\def\slash#1{\settowidth{\bredde}{$#1$}\ifmmode\,\raisebox{.15ex}{/}
\hspace*{-\bredde} #1\else$\,\raisebox{.15ex}{/}\hspace*{-\bredde} #1$\fi}
\newcommand{\mat}{\left ( \begin{array}{cc}}
\newcommand{\emat}{\end{array} \right )}
\newcommand{\matt}{\left ( \begin{array}{ccc}}
\newcommand{\ematt}{\end{array} \right )}
\newcommand{\matf}{\left ( \begin{array}{cccc}}
\newcommand{\ematf}{\end{array} \right )}
\newcommand{\vect}{\left ( \begin{array}{c}}
\newcommand{\evect}{\end{array} \right )}

\newcommand{\be}{\begin{eqnarray}}
\newcommand{\ee}{\end{eqnarray}}
\newcommand{\beq}{\begin{equation}}
\newcommand{\eeq}{\end{equation}}
\newcommand{\ba}{\begin{array}{ccc}}
\newcommand{\ea}{\end{array}}

\newcommand{\noi}{\vspace{12pt}\noindent}
\newcommand{\lG}{\raise.3ex\hbox{$\stackrel{\leftarrow}{G}$}}
\newcommand{\lU}{\raise.3ex\hbox{$\stackrel{\leftarrow}{U}$}}
\newcommand{\lP}{\raise.3ex\hbox{$\stackrel{\leftarrow}{{\cal P}}$}}
\newcommand{\leta}{\raise.3ex\hbox{$\stackrel{\leftarrow}{\eta}$}}
\newcommand{\lOmega}{\raise.3ex\hbox{$\stackrel{\leftarrow}{\Omega}$}}
\newcommand{\ldr}{\raise.3ex\hbox{$\stackrel{\leftarrow}{\delta^r}$}}

\def\beqn{\begin{eqnarray}}
\def\eeqn{\end{eqnarray}}

\def\gtwid{\raise.3ex\hbox{$>$\kern-.75em\lower1ex\hbox{$\sim$}}}
\def\ltwid{\raise.3ex\hbox{$<$\kern-.75em\lower1ex\hbox{$\sim$}}}

\def\t{{\mbox{\rm Tr}}}
\begin{document}
\topmargin -1.4cm
\oddsidemargin -0.8cm
\evensidemargin -0.8cm


\title{
{\vspace{-1.5cm} \normalsize
\hfill \parbox{40mm}{CERN-TH/2001-127}}\\
{\vspace{-0.35cm} \normalsize
\hfill \parbox{40mm}{hep-lat/0105010}}\\[25mm]
\Large{\bf Quenched Finite Volume Logarithms }}
\vspace{1.5cm}

\author{~\\{\sc P.H. Damgaard}\footnote{On leave from: The Niels
Bohr Institute, Blegdamsvej 17, DK-2100 Copenhagen, Denmark.}\\~\\
Theory Division\\CERN\\CH-1211 Geneva 23\\Switzerland}
\maketitle
\vfill
\begin{abstract}
Quenched chiral perturbation theory is used to compute the first finite
volume correction to the chiral condensate. The correction diverges
logarithmically with the four-volume $V$. We point out that with dynamical
quarks one can obtain both the chiral condensate and the pion decay
constant from the distributions of the lowest Dirac operator eigenvalues. 
\end{abstract}
\vfill

\thispagestyle{empty}
\newpage

\section{Introduction}

Finite-size effects in lattice gauge theory are usually considered 
undesireable features of numerical simulations. There are 
exceptions: near critical points, for instance, finite-size scaling
applies the Renormalization Group in finite volumes to extract 
critical indices of the infinite volume system. Also away from critical
points one can use knowledge of the finite volume behavior to learn
about the infinite volume theory. When this is the case, one is turning
the apparent disadvantage of being restricted to small volumes into
a distinct advantage. 

\noi
Ordinarily finite-volume computations in gauge theories still
involve the full complexity of such strongly coupled theories. A program
to overcome this obstacle was initiated several years ago by Gasser and 
Leutwyler \cite{GL1,GL2}. Let the finite (euclidean) 4-volume be of size
$V=L_xL_yL_zL_t(L_t=1/T)$. The idea is that for $L_i \gg 1/\Lambda_{QCD}$
the euclidean partition function (or appropriate generating functional)
is dominated by the lowest-mass excitations. If chiral symmetry is
spontaneously broken in the infinite volume limit
these light particles are the pseudo-Goldstone
bosons, whose interactions vanish at zero momentum transfer. The theory
governing these excitations is a chiral Lagrangian. Although
non-renormalizable, this theory can be treated perturbatively in the
sense of an effective Lagrangian. Combined with the description of
the fundamental degrees of freedom in the underlying gauge theories, this
allows for precise finite-volume computations in these otherwise
strongly coupled theories.  

\noi
There are two finite-volume regimes to consider. The first
is the one connected with the usual infinite volume limit of the theory.
If $M$ denotes the mass of the pseudo-Goldstone boson, one requires
$1/M \ll L_i$ so that the particle ``fits well inside the box'', and
the finite volume $V$ is a minor perturbation. In this regime usual
chiral perturbation theory is applicable. The other regime is   
$1/M \gg L_i$: if the volume is much larger than the scale set by
the gauge theory, $L_i \gg 1/\Lambda_{QCD}$, the euclidean partition
function is still dominated by the lightest excitations, but ordinary
chiral perturbation theory fails. In this region a different expansion
must be applied \cite{GL2}. This paper concerns  
(partially) quenched theory in the latter regime. We shall also comment
on how a known result from the unquenched theory can be used to extract
the pion decay constant $F$ in an unusual, but quite simple way. 

\noi
It is worthwhile stressing that although the finite volumes which are 
being probed here are unphysical in the sense that the pion correlation
length exceeds the boundaries of the finite volume, 
the observables one can
extract in this way are infinite-volume quantities. It is precisely
the purpose to use unphysical finite-volume expectation values to
extract these infinite-volume quantities directly, without any
numerical extrapolation to the infinite-volume limit.

\section{The Chiral Condensate}

We begin with a brief review of the Feynman rules for partially quenched
chiral perturbation theory based on the replica method \cite{DS}. Denoting
the number of ordinary physical (``sea'') quarks by $N_f$, one adds (in its
simplest form) $N_v$ ``valence'' quarks of common mass $m_v$
to theory, and takes 
the limit $N_v \to 0$ in the end. The starting point is then an effective 
theory based on the coset of 
$$
{\rm U}(N_f+N_v)\times {\rm U}(N_f+N_v) \to {\rm U}(N_f+N_v) ~,
$$ 
including the flavor singlet field. The Lagrangian can be parametrized
\beq
{\cal L} ~=~ \frac{F^2}{4}{\mbox{\rm Tr}}(\partial_{\mu}U(x)
\partial^{\mu}U^{\dagger}(x)) - 
\frac{\Sigma}{2}{\mbox{\rm Tr}}{\cal M}(U(x) + U^{\dagger}(x))
+ \frac{m_0^2}{2N_{c}}\Phi_0^2(x) + \frac{\alpha}{2N_{c}}\partial_{\mu}
\Phi_0(x)\partial^{\mu}\Phi_0(x) ~.
\label{Lchpt}
\eeq
when keeping only the usual leading terms in the small momentum expansion.
The field $
U(x) \equiv \exp[i\sqrt{2}\Phi(x)/F] \in$ U($N_f+N_v)$ lives
on the Goldstone manifold, and we have separated out the flavor singlet 
combination $\Phi_0(x) \equiv {\mbox{\rm Tr}}\Phi(x)$. For this reason
it is convenient to work in the quark basis \cite{BG} where $\Phi_{ij}
\sim \bar{\psi}_i\psi_j$. Let $M^2_{ij}\equiv (m_i+m_j)\Sigma / F^2$, and
\beq
{\cal F}(p^2)~\equiv~1+\frac{m_0^2+\alpha p^2}{N_c}\left(
  \frac{N_v}{p^2+M_{vv}^2}+\sum_{f=1}^{N_f} 
\frac{1}{p^2+M_{ff}^2}\right) ~.
\label{F}
\eeq
The momentum-space propagator for the off-diagonal mesons
$\Phi_{ij} \sim \bar{\psi}_i\psi_j, ~i \neq j$ is then read off to be
\beq
D_{ij}(p^2) ~=~ \frac{1}{p^2+M_{ij}^2} ~,
\eeq
while the corresponding propagator for the 
diagonal combination $\Phi_{ii} \sim \bar{\psi}_i\psi_i$ is
\beq
G_{ij}(p^2) ~=~ \frac{\delta_{ij}}{(p^2+M_{ii}^2)}-\frac{(m_0^2+\alpha
  p^2)/N_c}{(p^2+M_{ii}^2)(p^2+M_{jj}^2){\cal F}(p^2)} ~. 
\label{Gij}
\eeq
With these rules it is straightforward to take the partially quenched
limit of chiral perturbation theory: one simply sends $N_v \to 0$ at the
end of the calculation. It is simple to show \cite{DS} that it is equivalent
to partially quenched chiral perturbation theory based on the supersymmetric 
formulation \cite{BG,S,CP}, but we find the replica method easier to use.
It is straightforward to extend the method to a more elaborate chiral
Lagrangian involving higher order terms in the momentum expansion. Here
we restrict ourselves to the ${\cal O}(p^2)$ Lagrangian, supplemented
with the lowest order $\Phi_0$-terms (\ref{Lchpt}).

\noi
The conventional regime in which to apply (partially quenched) chiral
perturbation is the one where the four-volume $V$ well encloses
the pseudo-Goldstone bosons of masses $M_{ij}$, $i.e., ~L \gg 1/M_{ij}$. 
Although the four-volume $V$
then can be kept finite, the correlation length does not reach from boundary
to boundary, and the theory looks like a mildly perturbed infinite-volume
theory. In this regime chiral perturbation theory can be arranged as
a systematic expansion, roughly speaking in the small
momentum $p$. This is clearly an obvious regime of interest for lattice
gauge theory.

\noi
Actually, as pointed out first by Gasser and Leutwyler \cite{GL1},
another finite-volume regime can also be used to extract physical
observables. This is the extreme finite-volume regime where 
$L \ll 1/M_{ij}$. Although the pseudo-Goldstone bosons here apparently
violate the bound which one would normally impose on the finite-size
system, one can still perform analytical calculations here, 
and thus extract
physical quantities by comparison with Monte Carlo data. This 
regime is governed by a perturbative theory of the chiral Lagrangian which 
is sometimes called the $\epsilon$-expansion \cite{GL2} (not to be 
confused with an expansion away from critical dimensions). Let us first
introduce the mean length scale 
\beq
L ~\equiv~ V^{1/4} ~.
\eeq
The counting of orders of $\epsilon$ is defined by fixing $1/L = 
{\cal O}(\epsilon)$. The non-zero momentum modes thus count as
$p \sim {\cal O}(\epsilon)$ too, while $m_i \sim {\cal O}(\epsilon^4)$
(and hence $M_{ij} \sim {\cal O}(\epsilon^2)$;
$\mu_i \equiv m_i\Sigma V$ is of order 1). A systematic expansion
can be performed in which one is working to a given fixed order
in $\epsilon$ \cite{GL2}.

\noi
In the above regime the leading term of the chiral Lagrangian is given
by the zero momentum mode of the field $U(x)$. The method of
ref. \cite{GL2} treats this case by means of a 
collective field technique. To this aim, introduce a new variable $\xi(x)$ 
containing non-zero momentum modes only, and write
\beq
U(x) ~=~ ue^{i\sqrt{2}\xi(x)/F}u ~,
\label{xidef}
\eeq
where $u \in$ U$(N_f+N_v)$ is a constant collective field 
corresponding to the zero momentum modes of the pseudo-Goldstone bosons.
Let $U_0 \equiv u^2$, and introduce the flavor singlet
\beq
\Xi_0 ~\equiv~ \t\varphi_0 ~~,~~~~~~~{\mbox{\rm with}}~~~ U_0 ~\equiv~ 
\exp[i\sqrt{2}\varphi_0/F] ~.
\eeq
and $\Xi(x) \equiv \t\xi(x)$. The change of variables (\ref{xidef})
is next inserted into the path integral. The Jacobian is non-trivial,
but to lowest order it contains a term that only shift the vacuum energy,
without affecting the effective Lagrangian calculation below. We then find 
that the action to lowest order becomes
\beqn
\int\!dx~ {\cal L} &=& \int\!dx~\t\left[\frac{1}{2}
\partial_{\mu}\xi(x)\partial^{\mu}\xi(x)+ 
\frac{m_0^2}{2N_{c}}(\Xi(x)-\Xi_0)^2  + 
\frac{\alpha}{2N_{c}}\partial_{\mu}\Xi(x)\partial^{\mu}\Xi(x)\right]
\cr
&&- \frac{\Sigma}{2}\t\left[{\cal M}(U_0 + U_0^{\dagger})\left(V - 
\frac{1}{F^2}\int\!dx~\xi(x)^2\right)\right] ~.
\eeqn
It is convenient to add to this action a mass term for the field
$\xi(x)$, and only taking this mass to zero at the end of the calculations.
Because of the Gaussian damping term for the non-zero modes, 
$\xi(x)$ counts as being of order $\epsilon$. For the singlet field
$\Xi(x)$ the counting is not as straightforward. The kinetic energy
term for $\Xi(x)$ suppresses fluctuations in $\Xi(x)$ beyond 
${\cal O}(\epsilon)$, so that the mass term for $\Xi(x)$ is of
order $\epsilon(m_{0})^2\epsilon^{-2}$. The parameter $m_0$ is here
{\em a priori} free as far as counting in $\epsilon$ is concerned,
and we thus have exactly the same counting ambiguity as in
conventional partially quenched chiral perturbation theory
\cite{BG,S}: to any given order in the momentum expansion we should
in principle include all orders of $m_0^2$ at each order
in $M_{ij}^2$. What we need to know here 
is that we can take $m_0^2/(N_cF^2)$ to be parametrically small, 
and thus perform a simultaneous expansion in $\epsilon$ and
$m_0^2/(N_cF^2)$. This is what we will do below. 

\noi
Integrating out the fluctuation field $\xi(x)$ to one-loop order, we obtain
the one-loop improved effective Lagrangian for the zero momentum modes. We
are particularly
interested in the first correction to the chiral condensate $\Sigma$. 
What enters repeatedly in what follows is the pseudo-Goldstone boson 
propagator, evaluated at zero,
\beq
 \Delta(M_{ij}^2) ~ \equiv ~ \frac{1}{V}\sum_p\frac{1}{p^2+M_{ij}^2} ~,
\label{prop}
\eeq
and the corresponding quantity for the modes without zero momentum,
\beq
 \bar{\Delta}(M_{ij}^2) ~ \equiv ~ \frac{1}{V}\sum_p~\!'
\frac{1}{p^2+M_{ij}^2} ~. \label{prop'}
\eeq
The two are hence related by $\Delta(M_{ij}^2) = (M_{ij}^2V)^{-1} +  
\bar{\Delta}(M_{ij}^2)$. 

\noi
It is instructive to first rederive the known finite-volume correction for
the unquenched theory \cite{GL2} in the present formulation. We thus evaluate
the one-loop $\xi$-integral saturation of
$$
\left\langle\frac{m_s\Sigma}{2F^2}\t
\left[(U_0 + U_0^{\dagger})\int\!dx~\xi(x)^2
\right]\right\rangle_{\xi}
$$
where for convenience we have taken the $N_f$ dynamical fermions to be of 
equal mass $m_s$ (and the pseudo-Goldstone bosons thus have common masses 
$M_{ss} = 2m_s\Sigma/F^2$). Taking the $N_v=0$ expressions from above, we 
get\footnote{The coupling between $\Xi_0$ and $\Xi(x)$ is treated 
perturbatively. 
It contributes only to the next order.}
\beqn
\left\langle\int\! dx~\delta{\cal L}\right\rangle_{\xi} 
&=& \frac{Vm_s\Sigma}{2F^2}\t(U_0 + 
U_0^{\dagger})
\left\{(N_f-1)\bar{\Delta}(M_{ss}^2) + \frac{1}{V}\sum_p~\!' 
G(M_{ss}^2)\right\}
\cr
&=& \frac{Vm_s\Sigma}{2F^2}\t(U_0 + U_0^{\dagger})
\left\{(N_f-1)\bar{\Delta}(M_{ss}^2) + \bar{\Delta}(M_{ss}^2)
- \frac{1}{V}\sum_p~\!' \frac{\frac{1}{N_{c}}
(m_0^2+\alpha p^2)}{(p^2+M_{ss}^2)^2{\cal F}(p^2)}\right\} ~.
\label{unq1}
\eeqn 
This differs from the expression quoted in ref. \cite{GL2} because we have
here included the effect of the flavor singlet field. Due to the anomaly,
the mass of this field stays finite in the full theory as the
quark masses are decreased, while the pseudo-Goldstone bosons approach 
zero mass. This corresponds to taking $m_0^2 \to \infty$ in the above
expression, which after setting the argument of the propagator
to zero gives
\beq
\left\langle\int\! dx~\delta{\cal L}\right\rangle_{\xi} 
= \frac{Vm_s\Sigma}{2F^2}\t(U_0 + 
U_0^{\dagger})
\left\{\frac{N_f^2-1}{N_f}\bar{\Delta}(0)\right\} ~,
\label{fullcase}
\eeq
and we hence recover the result of ref. \cite{GL2}.

\noi
We now turn to the fully quenched case. This simply corresponds to setting 
$N_f=0$ in the Feynman rules given above. We evaluate
$$
\lim_{N_v\to 0}\frac{1}{N_v}\left\langle\frac{m_v\Sigma}{2F^2}\t
\left[(U_0 + U_0^{\dagger})\int\!dx~\xi(x)^2\right]
\right\rangle
$$ 
which is a finite quantity. In complete analogy with eq. (\ref{unq1}), this
results in
\beqn
&& \lim_{N_v\to 0}\frac{1}{N_v}\frac{Vm_v\Sigma}{2F^2}\t(U_0 + U_0^{\dagger})
\left\{(N_v-1)\bar{\Delta}(M_{vv}^2) + \frac{1}{V}\sum_p~\!'
G(M_{vv}^2)\right\}_{N_{v}\to 0}\cr
&=&  \lim_{N_v\to 0}\frac{1}{N_v}\frac{Vm_v\Sigma}{2F^2}\t(U_0 + U_0^{\dagger})
\left\{(N_v-1)\bar{\Delta}(M_{vv}^2) + \bar{\Delta}(M_{vv}^2)
- \frac{1}{V}\sum_p~\!' \frac{\frac{1}{N_{c}}
(m_0^2+\alpha p^2)}{(p^2+M_{vv}^2)^2{\cal F}(p^2)}\right\}_{N_{v}\to 0} \cr
&=& - \lim_{N_v\to 0}\frac{1}{N_v}\frac{Vm_v\Sigma}{2F^2}\t(U_0 + 
U_0^{\dagger})
\frac{1}{V}\sum_p~\!' \frac{\frac{1}{N_{c}}
(m_0^2+\alpha p^2)}{(p^2+M_{vv}^2)^2} ~,
\label{q1}
\eeqn
after doing the integration over the non-zero modes only.
As expected, this involves the quenched double pole \cite{BG,S}. 
We can rewrite the correction as
$$
- \lim_{N_v\to 0}\frac{1}{N_v}\frac{Vm_v\Sigma}{2N_{c}F^2}\t(U_0 + U_0^{\dagger})
\left\{\alpha\bar{\Delta}(M_{vv}^2) - (m_0^2 - \alpha M_{vv}^2)
\partial_{M_{vv}^{2}}\bar{\Delta}(M_{vv}^2)\right\} ~.
$$
Let us summarize what this correction implies. To zero'th order, the
mass-dependent quenched chiral condensate in finite volume $V$ is simply given
by a group integral. We introduce the partition function at zero
$\theta$-angle
\beq
{\cal Z}(\mu_v) ~=~ \int_{U(N_{v})}\!dU_0 \exp\left[\frac{1}{2}\mu_v
\t\left(U_0+U_0^{\dagger}\right) - \frac{m_0^2V}{2N_{c}}\Xi_0^2
\right] ~. \label{Z}
\eeq
It follows from the above analysis that
this is the zero'th order approximation of the effective euclidean
partition function of QCD with $N_v$ mass-degenerate (light) quarks in the 
finite-volume regime $V \gg
1/\Lambda_{QCD}^4$ and $M_{vv}^4 \ll 1/V$ \cite{LS}, including the effects
of a possibly light flavor singlet. Use has been made of the fact
that the first term in the effective Lagrangian depends only
on the combination $\mu_v \equiv m_vV\Sigma$.
Formally the quenched chiral condensate then follows from
\beq
\frac{\Sigma(\mu_v)}{\Sigma} ~\equiv~ \lim_{N_{v}\to 0}
\frac{1}{N_{v}}\frac{\partial}{\partial \mu_v}
\ln {\cal Z} ~. \label{qconddef}
\eeq
While it may be difficult to evaluate this expression explicitly, both
low-mass and large-mass series expansions have been obtained
in sectors of fixed topological gauge field charge $\nu$ \cite{DS1,DV}. 

\noi
To one-loop order in the fluctuations of the non-zero momentum modes
we see that this result is modified in a simple way. One can
use the same effective partition function (\ref{Z}) after replacing $\mu_v$ by
\beq
\mu_v' ~\equiv~ \mu_v\left\{1 + \frac{1}{N_{c}F^2}\left[
\alpha\bar{\Delta}(M_{vv}^2) - (m_0^2 - \alpha M_{vv}^2)
\partial_{M_{vv}^{2}}\bar{\Delta}(M_{vv}^2)\right]\right\} ~.\label{mushift}
\eeq
To the order shown, the mass-dependent chiral condensate still follows
from eq. (\ref{qconddef}), after making the above substitution in
the partition function (\ref{Z}), $viz.$,
\beq
\frac{\Sigma(\mu_v)}{\Sigma} ~\equiv~ \lim_{N_{v}\to 0}
\frac{1}{N_{v}}\frac{\partial}{\partial \mu_v'}
\ln {\cal Z}(\mu_v')\cdot\frac{\mu_v'}{\mu_v} ~. \label{qconddef'}
\eeq
This is completely analogous to the
case of dynamical quarks \cite{GL2}, where instead one should replace
the corresponding 
parameter $\mu_s$ by its modification due to (\ref{fullcase}).

\noi
Let us now compare this result with what one gets in ordinary quenched
chiral perturbation theory, which is valid in the opposite finite-volume 
regime where $V \gg 1/M_{vv}^4$. Then, to one-loop order \cite{BG,OTV} 
\beq
\frac{\Sigma(m_v)}{\Sigma} ~=~
1 + \frac{1}{N_{c}F^2}\left[\alpha\Delta(M_{vv}^2) - 
(m_0^2 - \alpha M_{vv}^2)
\partial_{M_{vv}^{2}}\Delta(M_{vv}^2)\right] + \ldots ~, \label{qs1loop}
\eeq
which also immediately follows from the quenched Feynman rules given above.
As in the case of dynamical fermions \cite{GL2} 
there is an overlapping
region where the two different expansions match on to each other.
To see this, let us evaluate the leading-order contribution to the
large-$\mu_v$ expansion based on the relations (\ref{Z}) and (\ref{qconddef}).
We thus need the asymptotic expansion of the partition function (\ref{Z})
for $N_v$ mass-degenerate fermions. To the best of our knowledge, such
an expansion has not been worked out yet. However, we can make use of
the known asymptotic expansions in sectors of fixed topological gauge
field charge $\nu$ \cite{DS1,DV}, and then perform the sum over topology
subsequently \cite{D}. We first note that the inclusion of an arbitrary
vacuum angle $\theta$ in the effective partition function (\ref{Z})
can be re-absorbed into a shifted $\Xi_0$, so that
\beq
{\cal Z}(\mu_v) ~=~ \int_{U(N_{v})}\!dU_0 \exp\left[\frac{1}{2}\mu_v
\t\left(U_0+U_0^{\dagger}\right) - \frac{m_0^2V}{2N_{c}}(\Xi_0-\theta F/
\sqrt{2})^2\right] ~. \label{Ztheta}
\eeq
The projection on sectors of fixed topological charge $\nu$ is
then performed by means of a Fourier transform \cite{LS}. The integral
is Gaussian, and the result is, for large four-volumes $V$,
\beqn
{\cal Z}_{\nu}(\mu_v) &=& \frac{1}{2\pi}\int_0^{2\pi}\! d\theta ~
e^{-i\nu\theta}{\cal Z}(\mu_v) \cr
&=& \frac{1}{\sqrt{\langle\nu^2\rangle}}e^{-\nu^2/2\langle\nu^2\rangle}
\int_{U(N_{v})}\! dU_0~ (\det U_0)^{\nu} \exp\left[\frac{1}{2}\mu_v
\t\left(U_0+U_0^{\dagger}\right)\right] ~,
\eeqn
up to an irrelevant $\nu$-independent normalization factor. We have 
then simultaneously identified the topological susceptibility \cite{OTV}
\beq
\langle \nu^2 \rangle ~\equiv~ \frac{F^2m_0^2V}{2N_{c}} ~. 
\label{tops}
\eeq

\noi
We now insert the large-$\mu_v$ series expansion of ${\cal Z}_{\nu}(\mu_v)$
(up to $\mu_v$-independent factors that are irrelevant right 
here) \cite{DS1,DV},
\beq
{\cal Z}_{\nu}(\mu_v) ~=~ \mu_v^{-N_v^2/2}e^{N_{v}\mu_v}\left(
1 - N_v\frac{\nu^2}{2\mu_{v}} + N_v(2N_v^2+1)\frac{1}{8\mu_{v}}
+ \ldots\right)
\eeq
into the definition of the quenched chiral condensate
\beq
\lim_{N_{v}\to 0}
\frac{1}{N_{v}}\frac{\partial}{\partial \mu_v}
\ln {\cal Z}(\mu_v) ~=~ 1 + \frac{\nu^2}{2\mu_v^2} -\frac{1}{8\mu_{v}^2}
+ \ldots 
~=~ 1 + \frac{\nu^2-1/4}{2\mu_{v}^2} + \ldots
\eeq
When next we wish to sum over topology \cite{D},
\beq
\frac{\Sigma(\mu_v)}{\Sigma} ~=~ \frac{\langle
\Sigma_{\nu}(\mu_v)\rangle}{\Sigma} ~=~
\frac{1}{\Sigma}\sum_{\nu=-\infty}^{\infty}{\cal Z}_{\nu}(\mu_v)
\Sigma_{\nu}(\mu_v)/{\cal Z}(\mu_v)
\eeq
we are faced with the fact that $\langle\nu^2\rangle \sim V$, while
$\mu_v$ here is kept finite as $V \to \infty$. This means that it
is not immediately obvious if there is a region in which we can expect to find
overlap between the two perturbative expansions discussed 
above.\footnote{Indeed, the leading-order term in the expansion, which
ordinarily is responsible for an overlapping region of validity 
\cite{GL2,OTV} of the two expansions is proportional to $N_v$, and hence 
vanishes in the quenched case. As shown below, in the quenched
theory the region of agreement is just pushed one order higher.}
The precise requirement for the $\nu$-average of the above first-order
term in the expansion to be small is
\beq
V ~\gg~ \frac{F^2m_0^2}{2N_{c}m_{v}^2\Sigma^2} ~,
\eeq
while we also need
\beq
V ~\ll~ \frac{1}{M_{vv}^4} ~=~ \frac{F^4}{4m_{v}^2\Sigma^2} ~.
\eeq
This can be achieved by having $m_0^2/N_c \ll F^2/2$, which formally
can be met by sending $N_{c} \to \infty$ while keeping all other quantities 
fixed\footnote{In a proper scaling with $N_c$, $m_0$ is of order 1,
while $F^2$ is of order $N_c$, but this only makes it more clear that
the inequality can be met.}. As discussed above, this limit is within
the region of validity of the finite-volume $\epsilon$-expansion,
and it gives a consistent regime in which we should
have simultaneous validity of both types of expansions. We can then
insert the relation (\ref{tops}) into the $\nu$-averaged condensate
\beq
\lim_{N_{v}\to 0}\left\langle
\frac{1}{N_{v}}\frac{\partial}{\partial \mu_v}
\ln {\cal Z}_{\nu}(\mu_v)\right\rangle ~~\simeq~~ 1 + 
\frac{\langle\nu^2\rangle}{2\mu_{v}^2} + \ldots ~~=~~
1 + \frac{m_0^2}{N_{c}F^2M^4V} + \ldots ~,
\eeq 
and hence, from eqs. (\ref{qconddef'}) and (\ref{mushift}), to leading order
\beqn
\frac{\Sigma(\mu_v)}{\Sigma} &=& \left[1 + \frac{m_0^2}{N_{c}F^2M^4V}\right]
\left\{1 + \frac{1}{N_{c}F^2}\left[
\alpha\bar{\Delta}(M_{vv}^2) - (m_0^2 - \alpha M_{vv}^2)
\partial_{M_{vv}^{2}}\bar{\Delta}(M_{vv}^2)\right]\right\} + \ldots \cr
&=&
1 + \frac{1}{N_{c}F^2}\left[
\alpha\Delta(M_{vv}^2) - (m_0^2 - \alpha M_{vv}^2)
\partial_{M_{vv}^{2}}\Delta(M_{vv}^2)\right] + \ldots ~.
\eeqn
Thus, in this particular regime the leading-order correction to
the chiral condensate precisely coincides with that of one-loop chiral
perturbation theory (\ref{qs1loop}). 

\noi 
An alternative way of interpreting the formulas (\ref{mushift}) and
(\ref{qconddef'}) is that, effectively, and to leading order in 
the finite-volume perturbation theory, one is working with a shifted 
quenched chiral condensate at finite volume:
\beq
\Sigma_{eff} ~=~ \Sigma\left\{1 + \frac{1}{N_{c}F^2}\left[
\alpha\bar{\Delta}(M_{vv}^2) - (m_0^2 - \alpha M_{vv}^2)
\partial_{M_{vv}^{2}}\bar{\Delta}(M_{vv}^2)\right]\right\} ~.
\label{Sigmaeff}
\eeq
It is clearly of interest to have this volume dependence made more
explicit. So far the calculation has been made at fixed ultraviolet
cut-off $\Lambda_{UV}$ (having in mind a lattice cut-off with
scale $\Lambda_{UV} \sim 1/a$). The explicit volume dependence is recovered
once one notices that in most regularization schemes the boson
propagator at the origin (\ref{prop'}), and any derivative thereof, 
separates into ultraviolet
divergent volume-independent terms, and additional finite 
volume-dependent terms \cite{HL}. The theory is renormalized as at infinite
volume, and the explicit volume dependence is what remains. A detailed
analysis has been made by Hasenfratz and Leutwyler \cite{HL}, from where
we can borrow the main results. In dimensional regularization, and setting
the subtraction point equal to unity, one has 
\beq
\Delta(M^2) ~=~ \frac{M^2}{16\pi^2}[\ln(M^2)+c_1] + g_1(M^2,L_i)
\eeq
where $c_1$ is a mass and $L_i$ independent constant that diverges near
four dimensions. We have ignored the tree level contribution from the
${\cal O}(p^4)$ terms in the chiral Lagrangian; they give volume-independent
terms that are of no interest here. All volume dependence is contained 
in the functions (in four dimensions; for $\Delta(M^2)$ we need only $r=1$) 
\cite{GL2,HL}
\beq
g_r(M^2,L_i) ~\equiv~
\frac{1}{V}\left(\frac{L^2}{4\pi}\right)^r \int_0^{\infty}\!
dt~t^{r-3}\exp\left[-\frac{M^2L^2t}{4\pi}\right]\left(
\prod_{i=1}^4 S(L_i^2/(tL))-1\right) ~,
\eeq
where
\beq
S(x) ~=~ \sum_{n=-\infty}^{\infty} \exp[-\pi n^2 x] ~.
\eeq
The general relation 
\beq
g_r(M^2,L_i) = -\frac{\partial}{\partial M^2} g_{r-1}(M^2,L_i)
\eeq
allows one to calculate all $g_r$'s from $g_0(M^2,L_i)$. For small masses
a systematic expansion has been given in ref. \cite{HL}:
\beq
g_0(M^2,L_i) ~=~ -\frac{2}{V}\ln(ML) + 
\frac{1}{16\pi^2}M^4\left[\ln(ML)-\frac{1}{4}\right] + 
\frac{1}{V}\sum_{n=0}^{\infty}
\frac{1}{n!}\beta_n(L_i/L) (ML)^{2n}
\eeq
where $\beta_n(L_i/L)$ are ``shape coefficients'' that depend on the shape
of the four-volume, and which can easily be evaluated numerically using
the formulas of ref. \cite{HL}. Subtracting the zero-momentum pole term,
we thus have 
\beqn
\bar{g}_1(M^2,L_i) &=& g_1(M^2,L_i) - \frac{1}{M^2V}\\ 
&=& \frac{1}{8\pi^2}M^2\ln(ML) - \sum_{n=1}^{\infty}
\frac{1}{(n-1)!}\beta_n(L_i/L)M^{2(n-1)}L^{2n-4} 
\eeqn
and
\beq
\bar{g}_2(M^2,L_i) = \frac{1}{8\pi^2}\ln(ML) + \frac{1}{16\pi^2}
+ \sum_{n=2}^{\infty}
\frac{1}{(n-2)!}\beta_n(L_i/L)M^{2(n-2)}L^{2n-4} ~, 
\eeq
from which we obtain both
\beq
\bar{\Delta}(M^2) ~=~ \frac{M^2}{8\pi^2}\left(c_1-\ln(L)\right) 
- \sum_{n=1}^{\infty}
\frac{1}{(n-1)!}\beta_n(L_i/L)M^{2(n-1)}L^{2n-4} \label{barDelexpl}     
\eeq
and its derivative
in a small-mass expansion. All volume dependence of the one-loop
result (\ref{Sigmaeff}) is thus explicitly given.

\noi
We note two important differences between the one-loop correction to
the quenched chiral condensate, and that of the full theory with
dynamical quarks:
\begin{itemize}
\item The expression for the effective finite-volume chiral condensate
(\ref{Sigmaeff}) involves two new constants ($m_0$ and $\alpha$), in
addition to the pion decay constant $F$.
\item In contrast to the case of dynamical quarks (\ref{fullcase}),
the leading correction contains not just the finite volume function
$g_1(M_{vv}^2,L_i)$, but also\footnote{The function $g_2(M_{ss}^2,L_i)$
appears at {\em next} order in the unquenched case, but then it is
suppressed by an additional factor of the volume $V$, and hence
harmless.}$g_2(M_{vv}^2,L_i)$.
\end{itemize}
The first point means that it is not straightforward to use the quenched
chiral condensate to extract the (quenched) pion decay constant $F$. The
second point implies, as we will see shortly, that the finite-volume
perturbative expansion of this quantity contains a logarithmic divergence.

\noi
We wish to find the effective chiral condensate at the finite 
volumes considered here. To avoid specifying the scale
entering the logarithm and all volume-independent terms, 
it is convenient to simply compare
$\Sigma_{eff}$ at two different four-volumes $V_1$ and $V_2$ of same
geometry (here, let $L_{1,2} \equiv V_{1,2}^{1/4}$). 
Based on (\ref{Sigmaeff}) and (\ref{barDelexpl}), we find 
\beq
\frac{\Sigma_{eff}(V_1)}{\Sigma_{eff}(V_2)} ~=~ 1 - \frac{1}
{N_{c}F^2}\left[\alpha\beta_1\left(
\frac{1}{L_1^2} - \frac{1}{L_2^2}\right) - \frac{m_0^2}{8\pi^2}\ln(L_1/L_2)
\right] + \ldots \label{S1/S2}
\eeq
after having taken the limit $M_{vv} \to 0$.
In contrast to the case of dynamical quarks, what should be a small
finite-volume correction can in fact become arbitrarily large. This
logarithmic divergence of the effective chiral condensate as the
volume $V$ is sent to infinity is a direct analog of the well-known
quenched chiral logarithms \cite{S,BG} of infinite-volume quenched
chiral perturbation theory. We call it a {\em quenched finite volume
logarithm}.

\noi
A logarithmic divergence in what should be a small correction of 
course signals that this finite-volume perturbative expansion of the quenched
effective theory is problematic. The trouble that
shows up as $\ln(M_{vv})$ for $M_{vv} \to 0$ in the regime 
$L \gg 1/M_{vv}$, shows up as $\ln(L)$ in the finite-volume regime
where $L \ll 1/M_{vv}$. Roughly speaking, as $1/M_{vv}$ is lowered and 
crosses $L$, the infrared cut-off switches from being given by $1/M_{vv}$ 
to being given by $L$. For both
expansions it is the very starting point of the expansion that appears
internally inconsistent. It seems that
at best one can hope that either of the two perturbative expansions have 
windows of validity where the one-loop corrections remain small. If
we adopt this point of view, then we should clearly demand in our case that
when comparing $\Sigma_{eff}$ at two different volumes $V_1$ and $V_2$
we must require
\beq
\frac{m_0^2}{8\pi^2N_{c}F^2}|\ln(L_1/L_2)| ~\ll~ 1 \label{logrange}
\eeq 
in order to apply (\ref{S1/S2}).
It has been argued by Sharpe \cite{S} that in usual chiral perturbation
theory it may be possible to resum a certain class of diagrams that give
rise to quenched chiral logarithms. However, also this partial
resummation remains divergent. Presumably a similar resummation may
be performed for the quenched finite volume logarithms. But we prefer
to avoid speculation beyond the one-loop order to which we have computed
the correction, and insist at least on the condition (\ref{logrange}).  
Difficulties with quenched chiral perturbation theory in finite volumes
(but associated
with unusual power-law behavior in $L$) have been noted earlier in
connection with the quenched pion scattering length \cite{BG1}.
 
\noi
A divergent chiral condensate for the quenched theory has long been
suspected from many different arguments, of which the presence
of ordinary (infinite-volume) quenched chiral logarithms \cite{S,BG}
is the one closest to the present perspective. 
The divergence has been difficult
to see in lattice gauge theory simulations, but a recent careful Monte Carlo
study of the quenched Schwinger model using overlap fermions has  
revealed a clear increase in $\Sigma$ as the volume $V$ is made larger
\cite{KN}. This
should be directly related to the quenched finite volume logarithm found
here, although the analysis will differ in detail due to the 
particularities of two space-time dimensions \cite{Sm,DSh}

\section{The Chiral Condensate and $F$ from Dirac Operator Eigenvalues} 

\noi
The dominance of the zero momentum mode in the effective chiral
Lagrangian when $1/M_{ii}^4 \gg V$ is known to have a very direct 
consequence in terms the Dirac operator spectrum of the underlying theory
\cite{LS,SV,ADMN}. In this finite-volume range the spectral properties
of the Dirac operator are analytically computable for an infinite
sequence of eigenvalues. While originally derived on the basis of
universal results from Random Matrix Theory \cite{SV,ADMN}, the 
low-energy spectrum of the Dirac operator
can equally well be obtained directly from the effective Lagrangian
framework \cite{OTV,DOTV,TV,Sz}. Let us denote 
the spectral density of the Dirac operator
by $\rho(\lambda;m_1,\ldots,m_{N_{f}})$, where $\lambda$ is
the eigenvalue. It was observed some time ago \cite{BBetal} that the
so-called microscopic spectral density \cite{SV},
\beq
\rho_s(\zeta;\mu_1,\ldots,\mu_{N_{f}}) ~\equiv~ 
\frac{1}{V\Sigma}\rho\left(\frac{\zeta}{V\sigma};\frac{\mu_{1}}{V\Sigma},
\ldots,\frac{\mu_{N_{f}}}{V\Sigma}\right) 
\eeq
gives an excellent finite-size scaling function from which to extract
the infinite-volume chiral condensate $\Sigma$ from finite lattice volumes.
Even more information can be obtained from the individual eigenvalue
distributions, which all depend on just this one single parameter
$\Sigma$, and whose complete analytical expressions are now
known \cite{NDW}. Each single eigenvalue distribution can then be used
to measure the infinite-volume chiral condensate $\Sigma$.

\noi
The preceding section clearly suggests that apart from the chiral condensate
one more quantity can be extracted from the microscopic Dirac operator
spectrum, namely the pion decay constant $F$. While the quenched case 
is special because of the quenched finite volume logarithm, the situation
is much more favorable when considering a theory with dynamical quarks.
We recall that the effective partition function, including its one-loop
correction (\ref{fullcase}), can be written
\beq
{\cal Z}_{eff} ~=~ \int_{SU(N_{f})}\! dU~  \exp\left[\frac{1}{2}
m\Sigma_{eff}V
\t\left(U_0+U_0^{\dagger}\right)\right] ~. \label{Zeff}
\eeq
where
\beq
\Sigma_{eff} ~=~ \Sigma\left[1 - \frac{N_f^2-1}{N_f}\frac{1}{F^2}
\bar{\Delta}(M_{ss}^2)
\right] ~. \label{Sigmashift}
\eeq
Taking the massless limit of eq. 
(\ref{Sigmashift}), the only correction to $\Sigma$ comes from
the finite four-volume:
\beq
\Sigma_{eff} ~=~ \Sigma\left[1 + \frac{N_f^2-1}{N_f}\frac{1}{F^2}
\frac{\beta_1(L_i/L)}{L^2} + {\cal O}(1/L^4)
\right] ~. \label{Sigmashift0}
\eeq
For symmetric four-tori $\beta_1 = 0.140461...$ \cite{HL}, and for other
shapes the coefficient $\beta_1$ is readily computed numerically from
the general formula of that same reference. The correction to $\Sigma$
is thus explicitly known to one-loop accuracy, and it depends only on $F$.

\noi
{}From the effective partition function (\ref{Zeff}), 
and its supersymmetric extension
\cite{OTV,DOTV,TV}, follows the microscopic Dirac operator spectrum. To
first order in perturbation theory it is unchanged in form compared with
the leading behavior, except for the finite-volume
shift indicated in (\ref{Sigmashift}).  It is immediately clear from 
(\ref{Zeff})
that all predictions for fixed sectors of topological charge $\nu$
are unchanged as well, except for the shift in $\Sigma$. With high enough 
statistics, the finite volume
correction to $\Sigma$ can be measured from fits to different
lattice volumes, yielding $F$. The two-loop contribution to the chiral
condensate in this volume range has also been computed \cite{Hansen},
but at that order one can no longer view the correction as just a
shift in $\Sigma$ in an otherwise unaffected effective theory.
The suggestion that the pion decay constant
can be extracted using finite-volume scaling in this form was first
made by Gasser and Leutwyler \cite{GL2}, and in a slightly different 
setting by Neuberger \cite{N}. The Dirac operator eigenvalues just give
a particularly convenient way to make use of the formalism.

\noi
The possibility of measuring the pion decay constant $F$ by considering
the combination of chiral perturbation theory and the analytical results for
the microscopic Dirac operator spectrum was first explored by 
Berbenni-Bitsch et al. in ref. \cite{BBetal1}. They compared lattice
Monte Carlo data for the chiral condensate in quenched SU(2) gauge theory
with a general expansion
of $\Sigma(m)$ for staggered fermions at finite four-volumes. 
Relating one of the coefficients
in the expansion to the pion decay constant $F$ by means of the
Gell-Mann--Oakes--Renner relation, they were thus able to extract $F$ from 
a general fit of $\Sigma(m)$ over a mass range of several orders of
magnitude. A similar analysis has very recently been made 
for staggered fermions and gauge group SU(3) \cite{Getal}. In that case
there are independent estimates of $F$ from earlier lattice studies, and
the authors of ref. \cite{Getal} find very good agreement.

\noi
A comment on staggered fermions can be made here. Comparisons between
the analytical prediction for the Dirac operator spectrum of dynamical
staggered fermions will typically be made so far from the continuum that 
the continuum number of flavors is not yet seen. Instead, the theory appears 
for just one staggered fermion to be of $N_f=1$ \cite{DHNR}, and
in the effective theory the remaining bosons are heavy
on the scale of the one single pseudo-Goldstone boson. The symmetry
breaking pattern is thus U(1)$\times$U(1)$\to$U(1), and it
is of interest to know also the leading correction for an unquenched
theory with this Abelian Goldstone manifold. Doing the one-loop 
finite-volume perturbative calculation we immediately find the analog of eq.
(\ref{fullcase}):
\beq
\left\langle\int\! dx~\delta{\cal L}\right\rangle_{\xi} 
= \frac{Vm_s\Sigma}{2F^2}(U_0 + 
U_0^{\dagger})
\bar{\Delta}(M_{ss}^2) ~,
\label{staggeredL}
\eeq  
where in this case $U_0 \in$ U(1).
The effective chiral condensate is then rescaled according to
\beq
\Sigma_{eff} ~=~ \Sigma\left[1 - \frac{1}{F^2}\bar{\Delta}(0)
\right] ~.\label{staggeredS}
\eeq
Thus also in this $N_f=1$ theory one can extract the pion decay constant
$F$ from the smallest Dirac operator eigenvalue distributions. The
general case of $N_f$ staggered fermions at strong coupling follows
similarly (a factor of $N_f$ in front of the correction term in 
(\ref{staggeredS})).

\noi
The above result for the $N_f=1$ theory with staggered fermions away
from the continuum limit may appear to contradict a recent lattice
Monte Carlo study of the distributions of the smallest Dirac operator
eigenvalues in SU(3) lattice gauge theory with one staggered fermion
\cite{DHNR}. There a very accurate agreement was found when comparing
with the analytical prediction based on the continuum $N_f=1$ theory,
where there is no Goldstone manifold at all, and all mesons are massive
even in the chiral limit. In fact, there is no contradiction.
The agreement found in \cite{DHNR} was with the sector of just topological
gauge field charge $\nu=0$. The projection on this sector is actually
equivalent to having a U(1) Goldstone manifold, as needed in order
to compare with the $N_f=1$ theory with staggered fermions away from
the continuum. In detail, the projection on the $\nu=0$ sector,
\beq
{\cal Z}_{\nu=0} ~~=~~ \frac{1}{2\pi}\int_0^{2\pi}\! d\theta
~{\cal Z}(\theta) 
~~=~~ \int_{U(1)}\!dU ~e^{\mu\cos(\theta)}
\eeq
is completely equivalent to the zero momentum mode effective partition
function for the coset of U(1)$\times$U(1)$\to$U(1) spontaneous
symmetry breaking, {\em without} any projection on the $\nu=0$ sector. 
The r\^{o}le of the zero-mode U(1) field is simply
being played by the angular integration that projects down
on $\nu=0$.
This also gives a complementary viewpoint to the
observation that staggered fermions seem insensitive to gauge field
topology at these couplings \cite{DHNR1} (see ref. \cite{FHL} for the
picture that emerges as the continuum limit is taken).

\section{Conclusions}

We have analyzed quenched chiral perturbation theory in the finite-volume
regime where the pseudo-Goldstone masses have Compton wavelengths much
larger than the linear extent of the volume, $1/M_{ss}^4 \gg V$. In this
regime a good starting point for perturbation theory is not the usual 
plane-wave solution of the fields. Rather, it is a collective variable
which is chosen to be the zero momentum mode of the given field. Using
this perturbation theory, and adding the (uniterated) vertex corresponding
to the singlet field $\Phi_0(x)$ as usual for the quenched theory,
we have computed the one-loop correction to
the quenched chiral condensate. It turns out that this correction 
becomes large when the volume is taken to infinity, in the chiral
limit as a logarithm in the volume $V$. This so-called quenched finite volume
logarithm prohibits taking the volume $V$ to infinity. Whether or not
it signals a true divergence in the quenched chiral condensate clearly cannot
be ascertained within this perturbative framework alone. Quenched
finite volume logarithms are likely to be a generic feature of quenched
chiral perturbation theory in the present finite-volume regime, and
should thus appear in other observables as well.

In the full theory with dynamical quarks the one-loop correction to the
chiral condensate in this finite-volume regime amounts to a
rescaling of $\Sigma$ in an otherwise unchanged effective theory. We
have used this fact to point out that the microscopic spectrum of the
Dirac operator, and in particular the individual smallest eigenvalue 
distributions, can be used to extract the pion decay constant $F$. This
entails determining numerically the leading correction to the well-known 
microscopic scaling for the Dirac operator spectrum near the origin,
and it appears to be feasible. It is quite different from the program
to extract these quantities and further couplings in the chiral
Lagrangian from (ratios of) matrix elements whose
expansions in partially quenched chiral perturbation theory are known 
\cite{S2,So}. So far the eigenvalue approach has only been tested away from
the continuum and only for the case of $\Sigma$. It would be
interesting to see if with present lattice volumes
the finite-volume corrections can
be determined to the accuracy required to extract $F$, and can be pursued 
towards the continuum.

\noi\noindent
{\bf Acknowledgments}\\
This work was supported in part by EU TMR grant no.
ERBFMRXCT97-0122. The author thanks J. Bijnens, C. Diamantini and
P. Hernandez for discussions.

\end{document}